\documentclass[sigconf]{aamas}

\usepackage{balance} 

\makeatletter
\gdef\@copyrightpermission{
  \begin{minipage}{0.2\columnwidth}
   \href{https://creativecommons.org/licenses/by/4.0/}{\includegraphics[width=0.90\textwidth]{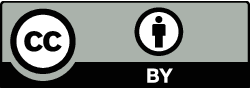}}
  \end{minipage}\hfill
  \begin{minipage}{0.8\columnwidth}
   \href{https://creativecommons.org/licenses/by/4.0/}{This work is licensed under a Creative Commons Attribution International 4.0 License.}
  \end{minipage}
  \vspace{5pt}
}
\makeatother

\setcopyright{ifaamas}
\acmConference[AAMAS '25]{Proc.\@ of the 24th International Conference
on Autonomous Agents and Multiagent Systems (AAMAS 2025)}{May 19 -- 23, 2025}
{Detroit, Michigan, USA}{Y.~Vorobeychik, S.~Das, A.~Nowé  (eds.)}
\copyrightyear{2025}
\acmYear{2025}
\acmDOI{}
\acmPrice{}
\acmISBN{}

\acmSubmissionID{1381}

\title[AAMAS-2025 Formatting Instructions]{Modeling the Centaur: Human-Machine Synergy in Sequential Decision Making}

\author{David Shoresh}
\affiliation{
  \institution{The Edmond and Lily Safra Center for Brain Sciences \\ Hebrew University}
  \city{Jerusalem}
  \country{Israel}}
\email{david.shoresh@mail.huji.ac.il}

\author{Yonatan Loewenstein}
\affiliation{
  \institution{The Edmond and Lily Safra Center for Brain Sciences \\
  Departments of Neurobiology and Cognitive Sciences
  and the Federmann Center for the Study of Rationality \\
  Hebrew University}
  \city{Jerusalem}
  \country{Israel}}
\email{}

\begin{abstract}
    The field of collective intelligence studies how teams can achieve better results than any of the team members alone.
    The special case of human-machine teams carries unique challenges in this regard. 
    For example, human teams often achieve synergy by communicating to discover their relative advantages, which is not an option if the team partner is an unexplainable deep neural network.
    Between 2005-2008 a set of "freestyle" chess tournaments were held, in which human-machine teams known as "centaurs", outperformed the best humans and best machines alone. 
    Centaur players reported that they identified relative advantages between themselves and their chess program, even though the program was superhuman.
    Inspired by this and leveraging recent open-source models, we study simulated human-machine teams in chess.
    A human behavioral clone ("Maia") and a pure self-play RL-trained chess engine ("Leela") were composed into a team using a Mixture of Experts (MoE) architecture.
    By directing our research question at the selection mechanism of the MoE, we could isolate the issue of extracting relative advantages without knowledge sharing. 
    We show that in principle, there is high potential for synergy between human and machine in a complex sequential decision environment such as chess.
    Furthermore, we show that an expert can identify only a small part of these relative advantages, and that the contribution of its subject matter expertise in doing so saturates quickly. 
    This is probably due to the "curse of knowledge" phenomenon.
    We also train a network to recognize relative advantages using reinforcement learning, without chess expertise, and it outdoes the expert. 
    Our experiments are repeated in asymmetric teams, in which identifying relative advantages is more challenging. 
    Our findings contribute to the study of collective intelligence and human-centric AI.
\end{abstract}


\newcommand{\BibTeX}{\rm B\kern-.05em{\sc i\kern-.025em b}\kern-.08em\TeX}

\begin{document}


\pagestyle{fancy}
\fancyhead{}


\maketitle 


\section{Introduction}

Collective intelligence studies how teams can outperform any one of their members.
The literature shows that in human teams, a moderate amount of diversity and open communication among members, are correlated with team synergy \cite{Malone}.
However, these conditions have limited applicability in human-machine teams, which have sharp cognitive diversity and which are communication-impaired, especially in the case of unexplainable deep neural networks.
We study the task of discovering relative advantages between team members in sequential decision making, and ask how it can be accomplished under these conditions.
In particular, we are interested in the role of domain expertise in making this judgment.

To answer these questions, we take inspiration from a set of eight "freestyle" chess tournaments held between 2005-2008 \cite{ChessBase}, in which participants could play either as a human, a chess program or as a "centaur".
In a centaur team, a human and machine work together analyzing the board positions, but the human makes the ultimate decisions.
The highest rank in these tournaments was consistently taken by a centaur team, despite the fact that the chess programs at the time were already superhuman. 
Centaur players have reported that knowing the relative advantages of their machine counterparts is an important element in their success \cite{ChessBase}.

This claim has an unintuitive implication: that the cognitive function of identifying relative advantages is not dependent on having superior domain expertise to the team members. 
\cite{Krakowski} have gone so far as to argue that in asymmetric human-machine teams, subject matter skills of the human aren't important at all, and that instead complementary human skills such as configuring the machine are the true sources of synergy.

To study this question, we performed computational experiments that leverage the recent availability of open-source chess models trained under different regimes:
a human-like behavioral clone trained with supervised learning from human games only, and a pure self-play RL agent with zero exposure to human data. 
The behavioral clone and the pure self-play RL agent are composed into a Mixture of Experts (MoE) model \cite{MoE}, which allows us to simulate human-machine teaming at scale, albeit in a simplified lab setting.
This also allows us to isolate the function of finding relative advantages, in the selection mechanism of the MoE.

We compare different selection mechanisms for the MoE: an oracle, a chess expert and an RL-trained team manager.


\section{Experimental Setup}

\subsection{Team Players}
We deploy two types of chess models to play as a team. 
One is Leela-Chess-Zero ("Leela"), an open-source version of Alpha-Zero \cite{Silver}, which is a pure self-play reinforcement learning algorithm. 
Due to its lack of exposure to human data or knowledge, it has been noted for its non-human or "alien" style of play \cite{Manella}.
The other is Maia-Chess ("Maia"), an open-source family of models designed to be human-like, trained exclusively from human games \cite{McIlroy}. 
Maia and Leela play as a team against a popular chess engine known as Stockfish, which has a heuristics-based evaluation function, parameterized by a shallow neural network.

There are multiple Maia, Leela and Stockfish versions available online. 
In all our experiments we used a fixed Maia network ranked 1900 ELO (a strong pre-master human ranking in chess).
Maia is deployed without search (depth=1) in order to maintain its human likeness.
We experiment with different levels of Leela and Stockfish strength levels compared to Maia. 
Leela is also deployed without search, and it's strength is increased by taking stronger networks.
Stockfish strength is adjusted using version and search depth.

\subsection{Team Model}
Maia plays in a MoE team with a Leela network against Stockfish.
The protocol for team play is as follows.
At each board position \emph{s} arrived at during a game, if Maia and Leela agree on the best move, that move is played. 
If Maia and Leela disagree on the best move, a manager chooses between their recommendations. 
If the manager is indifferent, the move is chosen randomly from the two recommendations.
Moves are chosen by this protocol from the start state until the end of the game.

This setup can be formally described as a Markov Decision Process (MDP)\cite{Sutton}, consisting of states \emph{S}, actions \emph{A}, a transition function \emph{T}, a reward function \emph{R}, and a discount factor $\gamma$ giving the tuple (\emph{S}, \emph{A}, \emph{T}, \emph{R}, $\gamma$).
A policy $\pi$ chooses actions given states. 
Expected accumulated reward of a policy, starting from a state-action pair, is known as the \emph{Q} function of the policy.

In our case, \emph{T} is the adversary, \emph{R} is the outcome of the game (1 for win, 0.5 for draw, 0 for loss) and $\gamma=1$.

Definition 1: to formalize a "team", we define a MoE policy, $\hat{\pi}$, that consists of a set of base policies ($\pi_1, \pi_2$), i.e. team members, and a selection mechanism between team members at each state which we denote the "manager".
We consider two types of managers, one that decides based on the state and the move recommendations ($\text{Manager}(s, a_{1}, a_{2})\rightarrow k\epsilon\left[1,2\right]$), and one that decides only based on the state ($\text{Manager}(s)\rightarrow k\epsilon\left[1,2\right]$). 

\[
  {\hat{\pi}}_{\left(\pi_1,\pi_2, \text{Manager}\right)}\left(s\right)=\pi_{k}\left(s\right)
\]

Definition 2: we seek a manager that achieves "synergy", i.e. its team has higher expected reward than any of the individual team members, in expectation over the initial states of the environment (in our case, chess opening positions).

\[
  \mathbb{E}_S(Q_{{\hat{\pi}}_{\left(\pi_1, \pi_2, \text{Manager}\right)}}(s,a_{{\hat{\pi}}_{\left(\pi_1,\pi_2, \text{Manager}\right)}}))>
\]
\[
\text{max}\left[\mathbb{E}_S(Q_{\pi_1}\left(s,a_{\pi_1}\right)), \mathbb{E}_S(Q_{\pi_2}\left(s,a_{\pi_2}\right))\right]
\]

where \emph{S} is the set of start states and \emph{a} are actions recommended by the subscripted policy in the corresponding state.

In this model, the original human centaur player is conceived as having two distinct functions: forming a substantive opinion about the best action (as one of the base policies) and estimating relative advantages (as the manager).
In other settings, one could also envisage an automated manager that gives a relative advantage score to team members in each state.

\subsection{RL Trained Manager}

\begin{figure}[h]
    \centering
    \includegraphics[width=1.0\linewidth]{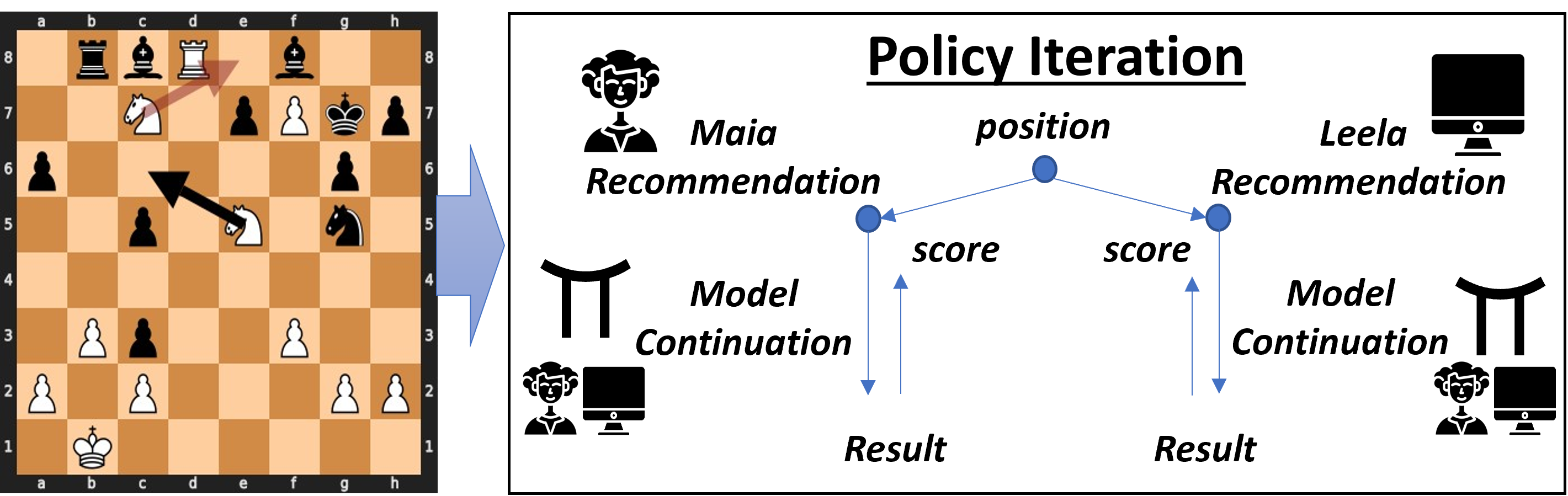}
    \caption{Process for obtaining estimated \emph{Q} values for RL}
    \label{PolicyIteration}
    \Description{Process for obtaining estimated \emph{Q} values}
\end{figure}

We trained a manager without independent knowledge of chess to identify relative advantages of the team players with RL, using game outcomes as reward. 
To train this manager, we used the policy iteration algorithm. 
At each iteration \emph{i} of the algorithm, we played a set of games, recording all disagreements between Maia and Leela. 
From each disagreement, Maia and Leela recommended moves were rolled out separately, with the rest of the game controlled by the current RL manager. 
The outcomes from these rollouts serve as the reward (see figure \ref{PolicyIteration}). 
For each decision point this give the tuple:

\[
  \{s,Q_{{\hat{\pi}}_{\left(\pi_M,\pi_L, \text{RLManager}_{i}\right)}}\left(s,\ a_M\right),\ Q_{{\hat{\pi}}_{\left(\pi_M,\pi_L, \text{RLManager}_{i}\right)}}\left(s,\ a_L\right)\}  
\]

where \emph{s} is the position being evaluated, $a_{M}$ and $a_{L}$ are the move recommendations in \emph{s} and $\text{RLManager}_{i}$ is the RL trained manager at iteration \emph{i}.
$Q_{\hat{\pi}}$ represents the \emph{Q} value for the team given the action recommendations $a_{M}$ and $a_{L}$, estimated empirically by the outcome of the game rollouts from \emph{s}. 
The resulting dataset was then used to train $\text{RLManager}_{i+1}$. 

\begin{figure}
    \centering
    \includegraphics[width=1.0\linewidth]{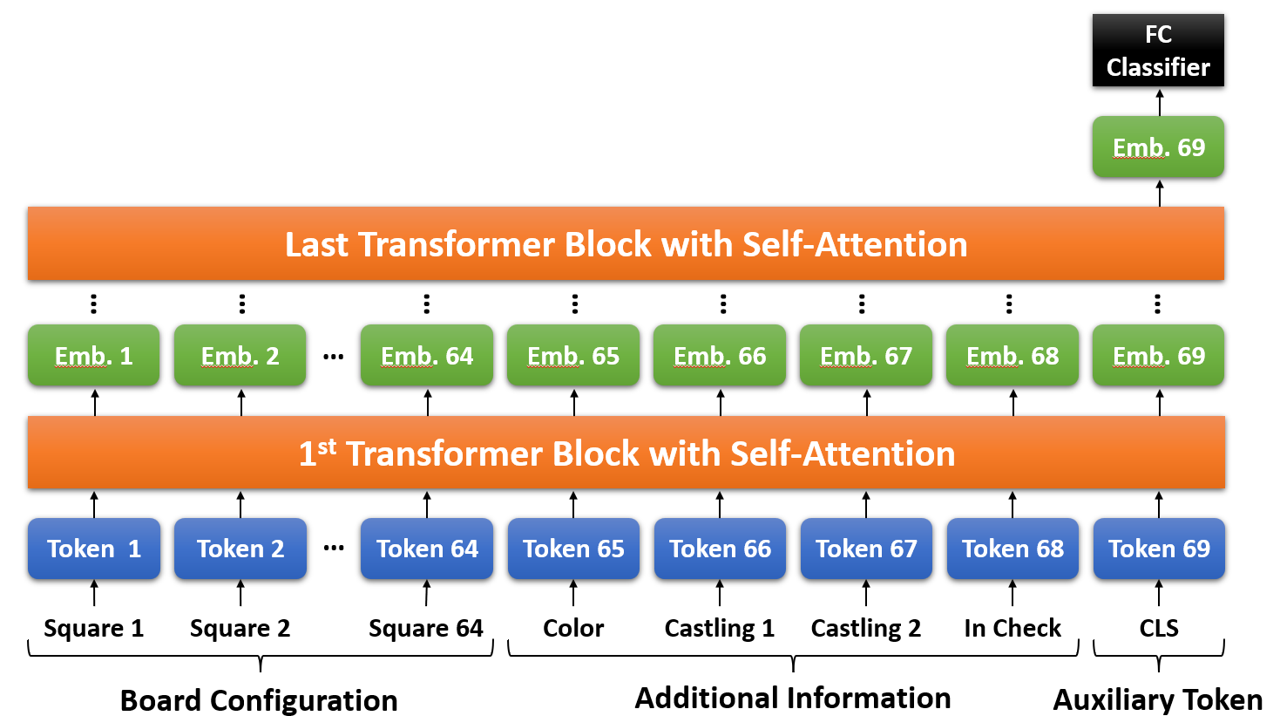}
    \caption{Transformer model architecture}
    \label{architecture}
    \Description{Transformer model architecture for chess manager}
\end{figure}

The RL manager was parameterized by a transformer architecture (illustration in figure \ref{architecture}). 
As input, the model receives a representation of the board using 64 tokens, one for each square. 
The tokens denote the pieces that occupy each square, or if it is empty, plus positional encoding. 
Additional tokens signify the color being played, castling rights and whether the king is in check. 
Following BERT \cite{Devlin}, there is also an auxiliary "CLS" token. 
The CLS token in the last layer is fed to a fully-connected classifier network which outputs a binary indicator for Maia or Leela.

Note that the RL manager receives only the state as input, without move recommendations. This is to train a manager focused on identifying the relative advantages of the team members, rather than judging the merits of the particular recommendations which would require deep domain expertise.

Chess engines are often developed and evaluated on games starting from diverse opening positions, to increase coverage of the game space. 
We collected over 28000 chess opening positions for training games.


\subsection{Expert Manager}

To examine the role of domain expertise in producing team synergy, we use an external strong "subject matter expert" chess engine as the manager. 
The expert has access to the recommended moves of Maia and Leela, and it scores out each of them using its own evaluation function. 
Formally, if we define a subject matter expert policy $\pi_{SME}$, we can define an expert-controlled team as:

\[
{\hat{\pi}}_{\left(\pi_{SME},\pi_M,\pi_L\right)}\left(s\right)=\begin{matrix}\text{Argmax}\\(a_M,a_L)\\\end{matrix}\left[Q_{\pi_{SME}}\left(s,\ a_M\right),Q_{\pi_{SME}}\left(s,\ a_L\right)\right]  
\]

Note also that in a non-sequential decision task, such an expert is the best possible manager.
In a sequential decision task this is less clear.
This is because the assumption when evaluating $Q_{\pi_{SME}}$ is that the subject matter expert $\pi_{SME}$ continues alone after the chosen move, which is not what happens in the team.
For the expert manager, we used a strong version of Stockfish (v. 14), experimenting with different search depths.

\subsection{Oracle Manager}

To compare the performance of these managers to the maximal potential synergy in a given Maia-Leela pair,
we can posit an additional "oracle" manager, that has online access to the team player and adversary policies. At each decision, it opens out full game trees, obtains each player's future move recommendations and adversary responses, and backward induces the best current recommendation.
This would not be an admissible manager during games, since it has full access to the adversary policy for the simulations.
But it would theoretically represent the highest attainable synergy within the team, allowing us to see how much the other managers fall short.

Practically, such an oracle is computationally demanding. 
As an alternative, we conceptualize an approximate oracle that simulates each of the recommended moves, and thereafter continues these two simulated branches with just one trajectory each, generated by continuing with the better of the two team players against the fully modeled adversary.
This approximate oracle then returns to the original decision point and chooses the player whose recommended move led to the better outcome in the simulation.
Since this only approximates the full theoretical oracle, it will not necessarily produce the optimal team performance, but it would provide a lower bound for that optimum.

\subsection{Evaluation}
We collected a random set of 500 opening chess positions (different from the ones used in training the RL manager), playing both black and white perspectives for each position, giving 1000 games in total.
To rate performance, we use the “Wins-Draws-Losses” score (WDL) defined as:

\[
  \text{WDL}=\frac{\text{\#wins}+\frac{1}{2}\cdot \text{\#draws}}{\text{\#wins}+\text{\#draws}+\text{\#losses}}
\]
\newline
The baseline for synergy is the WDL of the best team member playing by itself against the adversary: max($\text{WDL}_{\pi_{M}}, \text{WDL}_{\pi_{L}}$).

\section{Results}
\subsection{Results for a Symmetric Team}

\begin{figure}
    \centering
    \includegraphics[width=1.0\linewidth]{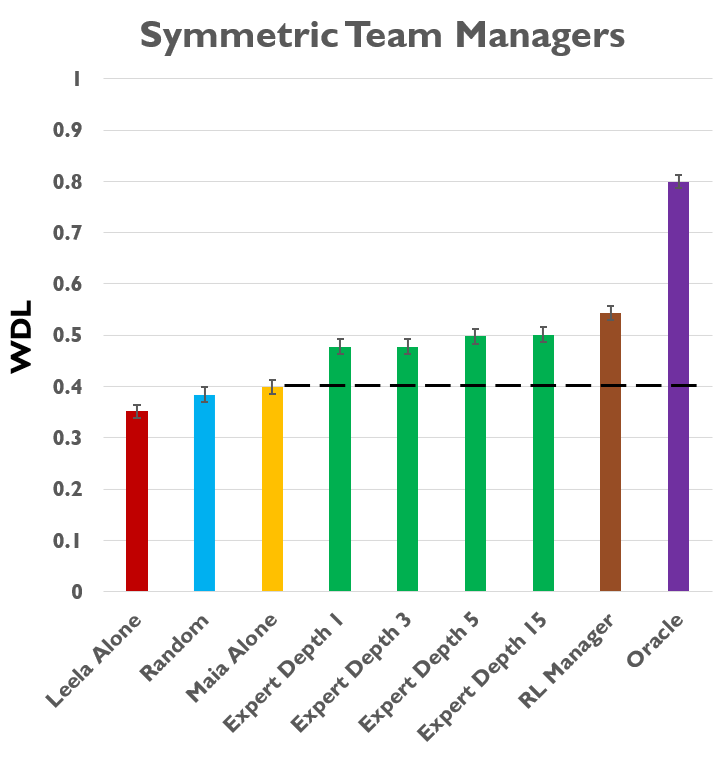}
    \caption{WDL of approximately symmetric team using different managers. "Random" refers to a random mixture policy with p=0.5. For the subject matter expert, we display results for a Stockfish engine manager set to depths 1, 3, 5 and 15. Synergy is above the line of WDL for Maia alone. Error bars represent SEM.}
    \label{SymmetricResultsFigure}
    \Description{Results for approximately symmetric centaur team using different managers}
\end{figure}

First we examine a symmetric team, i.e. where the Leela network is almost the same strength as the Maia network, playing against Stockfish (v. 11) with search depth=1.
Results are seen in figure \ref{SymmetricResultsFigure}.

Maia and Leela WDL scores when playing against Stockfish alone were 0.3995 and 0.3515 respectively.
One might think that a random mixture policy of the two players would be worse than either player alone, since the shifts between them would prevent coherent plans, or alternatively it might achieve synergy by introducing diversity.
In practice, a random mixture policy with p=0.5 produces a WDL in between that of the two players alone.

For the subject matter expert manager, we deployed another Stockfish engine (v. 14) with search depth=1 to score the recommended moves.
This team achieves significant synergy (WDL=0.4775) compared to Maia alone ($Z=7.23$).

To more closely examine the role of subject matter expertise, the manager was then deployed at increasing levels of search depth (3, 5 and 15).
At depth 15, if the expert was playing against the adversary itself without being constrained to Maia or Leela recommendations, it would have WDL score of 0.942.
All versions of the subject matter expert achieve significant team synergy, but increasing the manager's strength does not yield much improvement (team WDL at peak was 0.501). 
This seems to indicate that the marginal contribution of domain knowledge to finding relative advantages saturates quickly.
The limitations of the expert as manager can be viewed as a demonstration of the "curse of knowledge" \cite{Curse}.
Intuitively, if the manager is too smart, it may not properly account for the future behavior of the inferior team players after the current move.

The RL manager, which was not explicitly trained to play chess and has no access to the recommended moves, achieves WDL=0.5435, which is significantly better than the subject matter expert team ($Z=2.95$).
This gives an initial indication that to choose the best recommendations, it is more important to know the qualities of the base players than it is to be a subject matter expert.

The oracle achieved WDL of 0.799 far outperforming the other managers, indicating vast headroom for synergy.

\newpage
\subsection{Results for Asymmetric Teams}

We repeated the above experiment for three different levels of asymmetry within the Maia-Leela team.
The Maia network was held fixed, while successively stronger Leela networks were chosen to team with it.
The Stockfish adversary (v. 14) was also successively improved by setting it's search depth to levels 4, 7 and 10 respectively.
Leela and Stockfish were calibrated to be approximately on par with each other in each experiment, that is Leela has a WDL consistently close to 0.5.
The subject matter expert used in all teams was Stockfish (v. 14) at search depth=15.

\begin{figure}
    \centering
    \includegraphics[width=1.0\linewidth]{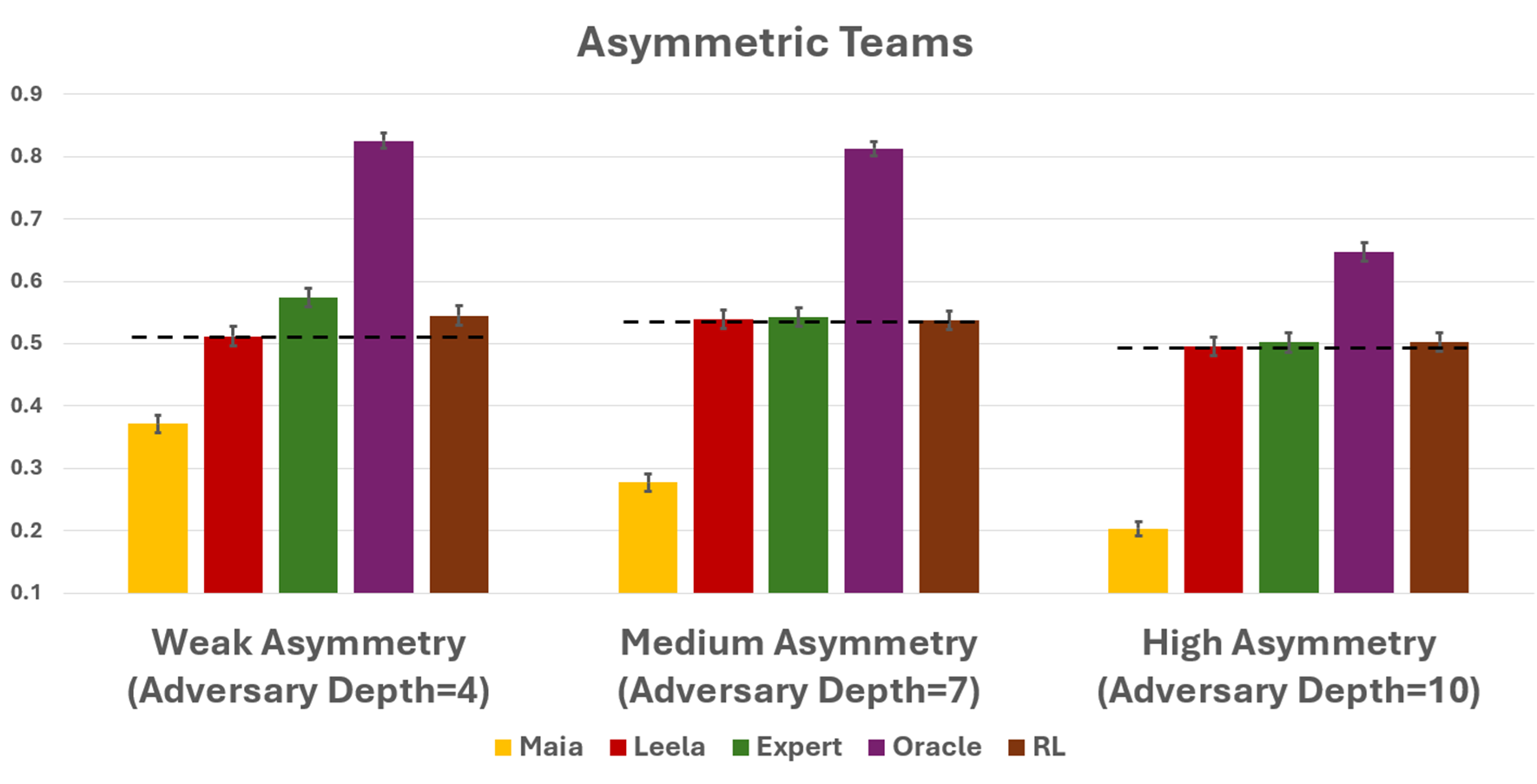}
    \caption{WDL for asymmetric teams. Synergy is above the WDL of Leela alone (dashed line). Error bars represent SEM.}
    \label{Asymmetric_Comparison}
    \Description{Results for asymmetric teams using different managers}
\end{figure}

The histogram in figure \ref{Asymmetric_Comparison} summarizes the results across the different levels of team asymmetry.
Perhaps as expected, all managers perform less well as asymmetry in the team increases.
In the first level of asymmetry, the RL manager (with comparable amount of training to the symmetric team RL manager) achieves statistically significant synergy compared to Leela alone ($Z=2.18$), but less than that of the subject matter expert.
Neither the subject matter expert nor the RL manager produce synergy in the more asymmetric teams.
Impressively though, the oracle shows that considerable potential synergy persists even in highly asymmetric teams.

\subsection{Distribution of Team Member Choices}

We also checked the proportion of states in which the managers assign decisions to Maia and Leela.
When the manager is indifferent, the choice between Maia and Leela is made randomly, so these instances are discounted.

\begin{figure}
    \centering
    \includegraphics[width=1.0\linewidth]{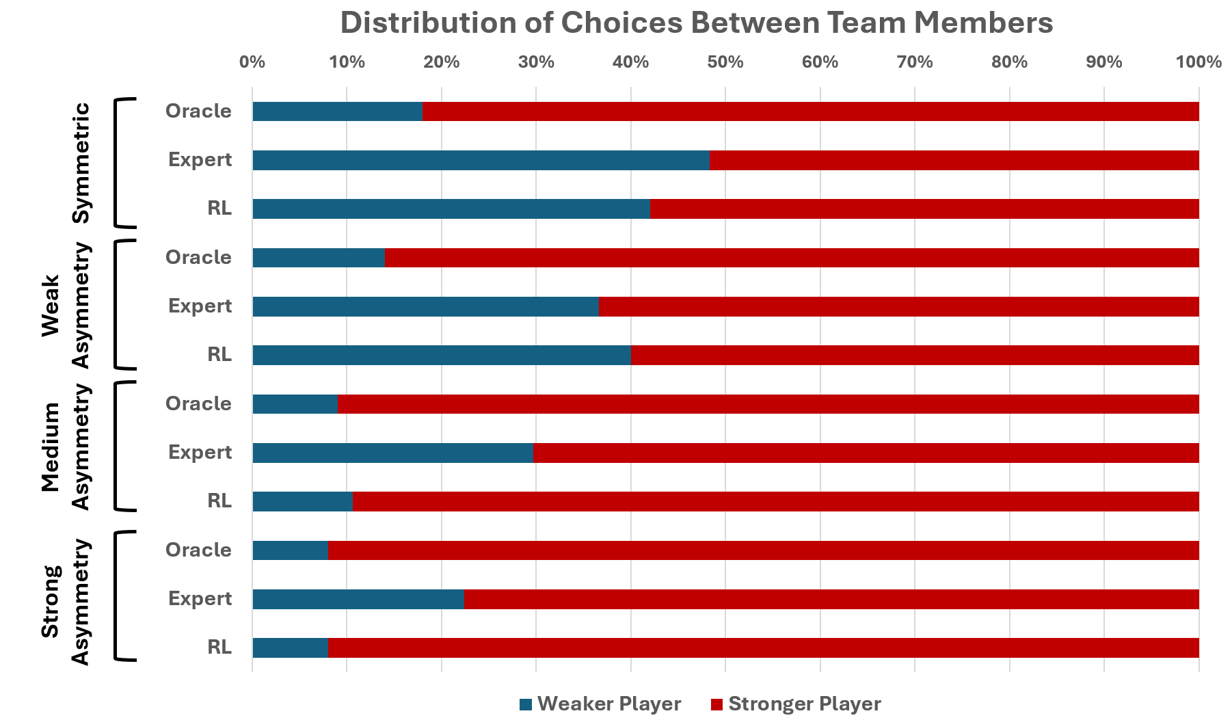}
    \caption{Distribution of choices between Maia and Leela. The oracle chooses the inferior player (Leela in the symmetric team and Maia in the others) more sparingly than the other managers.}
    \label{ProportionChoices}
    \Description{Distribution of choices between Maia and Leela by different managers in different teams}
\end{figure}

Results are show in figure \ref{ProportionChoices}.
Focusing on the oracle, it tends to choose the inferior player quite sparingly.
This may be due to the bias introduced by simulating the continuation after each move recommendation using the superior team member.
Nonetheless, when considering the high performance of the oracle, it seems to have identified the most consequential contributions by the inferior player.
We can infer that there is a small kernel of high-impact decisions, which makes it apparent why finding them is hard for the other managers.

\section{Explaining the RL Manager}

Recall that we trained the RL manager without exposing it to any rules of chess or to the recommended moves of the team players.
It was only fed static positions and binary Maia/Leela labels.
We compared it to the expert manager since we wanted to know if domain knowledge is decisive in the task of finding relative advantages. 
However, it is possible that the RL manager developed chess knowledge implicitly, as well as knowledge of the team members' policies, and uses these to identify the best recommendation. 
That is, it might have learned to become a subject matter expert.

In the following, we address the question of how the RL manager identifies relative advantages, by investigating its network.
For this analysis we use the manager trained for the symmetric team.

\subsection{Is domain knowledge necessary to identify relative advantage?}

First, we ask whether the RL manager uses non-trivial understanding of the domain to make its decisions.  
We explore this using the attention scores in the transformer network. 
Attention scores are used in transformer models to weight input tokens (in our case squares on the chess board) during processing, which is sometimes leveraged for explainability \cite{Vaswani, Wiegreffe}.

To see whether the RL manager learns something about the rules of chess, we checked if the network gives more attention to pieces than to empty squares. 
Over 8000 chess positions were fed to the network.
Attention scores of the CLS token towards tokens representing squares on the board were recorded (averaged over layers and attention heads).
For each position the mean attention to pieces and to empty squares were taken. 
For control, we compared to the same attentions given by an untrained network. 
In the untrained network, attentions were near uniform.
In the trained network, there was a clear difference in attention to pieces and to empty squares, to the point that the two distributions hardly overlap.

\begin{figure}
  \centering
  \includegraphics[width=1.0\linewidth]{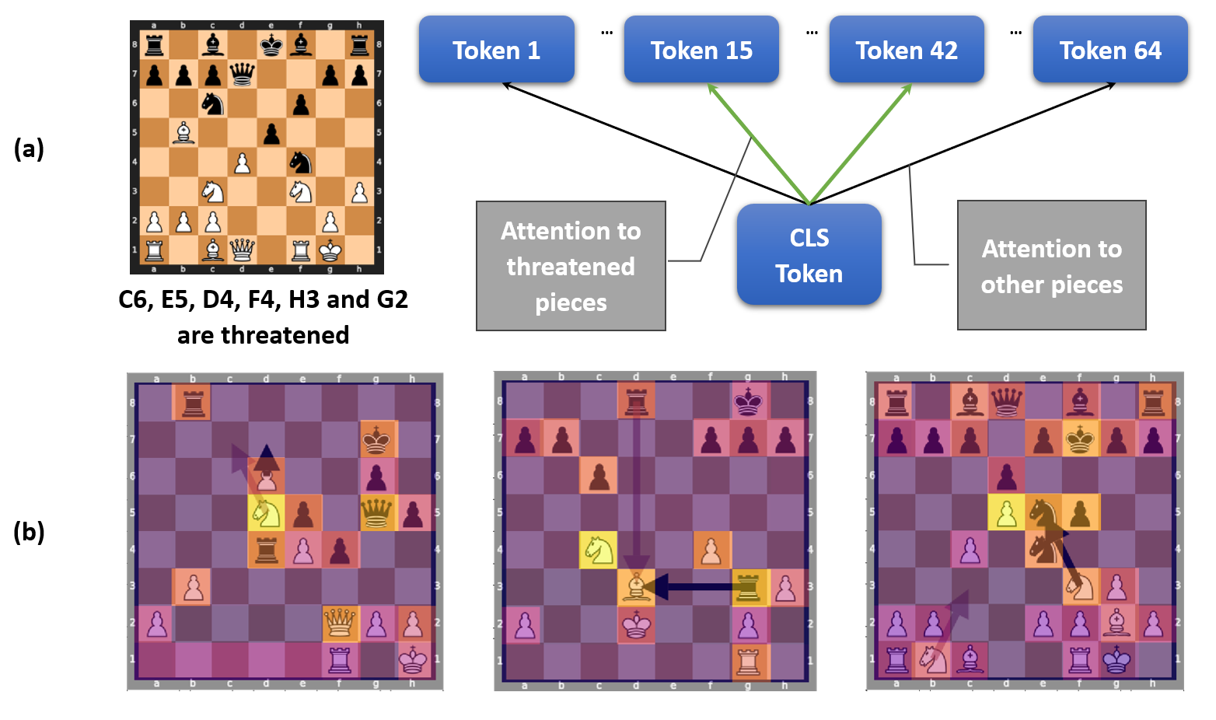}
  \includegraphics[width=1.0\linewidth]{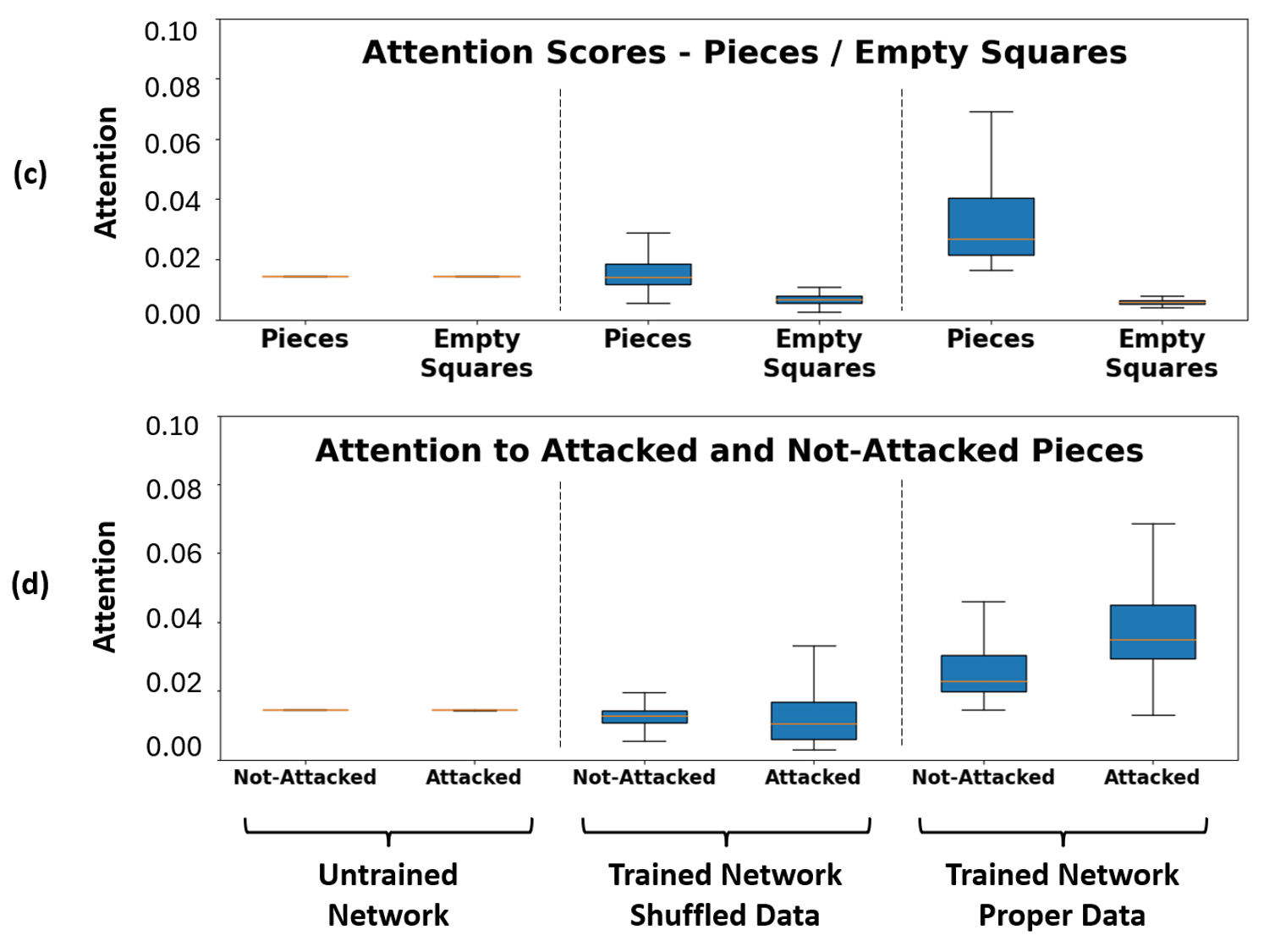}
  \caption{Investigation of the RL-manager network. (a) Illustration of attentions extracted from the model; 
  (b) Example attention heatmap overlay (lighter colors denote higher attention);
  (c) Attention to pieces vs empty squares;
  (d) Attention to attacked pieces vs not-attacked.
  Whiskers in the box plots represent farthest data point lying within 1.5x the inter-quartile range from the box.
  }
  \label{DomainKnowledge}
    \Description{Investigation of attention scores under rules of chess}
\end{figure}

Since pieces are correlated with regions of the board, we used an additional control - the trained network attentions over shuffled chess positions. 
In this control, the network gives more attention on average to pieces than empty squares, but not as pronounced as in the properly ordered data. 
Results are shown in figure \ref{DomainKnowledge}(c).

For a more advanced test, we checked if the network gives more attention to pieces that are attacked, than those that aren't, even though the network was never shown how pieces move in the game. 
Once again, we controlled with attention in an untrained network, and attention in the trained network over shuffled positions.

Results are shown in figure \ref{DomainKnowledge}(d).
As before, the untrained network gives near-uniform attention to all squares. 
With the shuffled data, there is no preference for attacked pieces over non-attacked. 
But over the ordered data the trained network clearly gives more attention to attacked pieces. 
We measure effect size using non-parametric $A_w$ which is a robust test for group differences in numeric variables \cite{Li}. 
It gives a high effect size of 0.81 (1 is highest).

The network focuses more on pieces than empty squares, and more on threatened pieces than non-threatened. This indicates that the manager obtained some domain knowledge while learning Maia and Leela relative advantages. 
We don't know how deep the domain knowledge goes, but at least, the model locates the pieces and observes their interrelationships under the rules of chess.

\subsection{Does the RL manager implicitly predict recommendations of team members?}

If the RL manager learned some chess, did it inadvertently become an expert manager?
For this to happen, it would have to learn to predict the move recommendations, since it is not given them.

\begin{figure}
  \centering
  \includegraphics[width=1.0\linewidth]{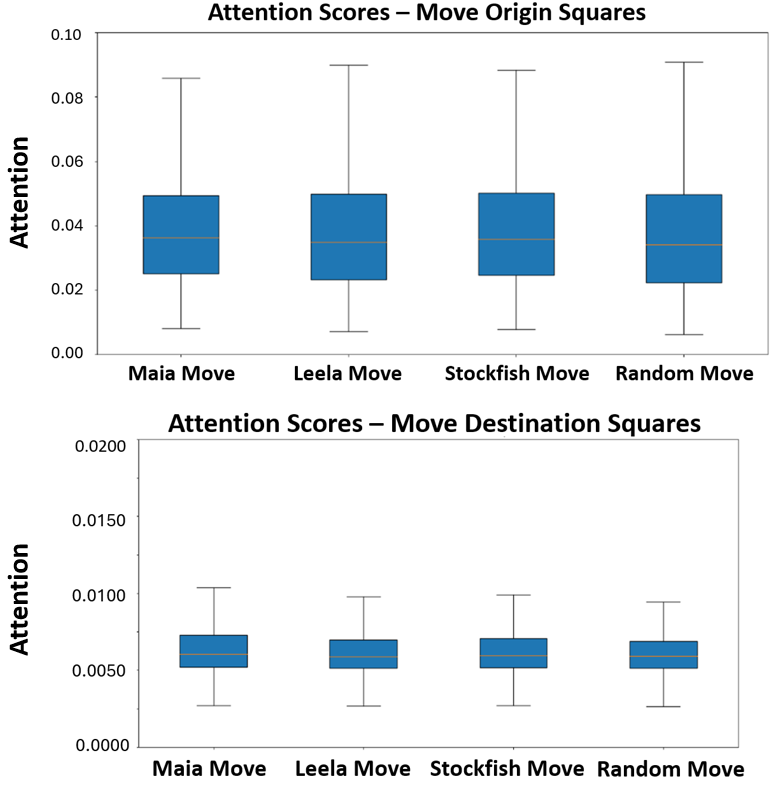}
  \caption{Comparison of attention to moves. Top - attention to origin squares, Bottom - attention to destination squares.
  No major difference is observed compared to the controls.
  Whiskers in the box plots represent farthest data point lying within 1.5x the inter-quartile range from the box.}
  \label{boxplots}
  \Description{Investigation of attention scores to moves by different players}
\end{figure}

To test for this, we ascertain whether the CLS token in the network gives more attention to Maia or Leela recommended moves than to other legal moves. 
Since moves are comprised of two squares, the origin and the destination, we measured the attention to each. 
We used two control variables - recommendations by a third-party chess engine (Stockfish, depth=5) and random legal moves. 
10000 board positions were taken for this test. 
As before, attention scores were averaged over layers and attention heads. 
Results are shown in figure \ref{boxplots}.
Maia and Leela move-recommendations do not seem to receive higher attention than the controls.

We thus find no evidence that the network learns to recognize Maia or Leela moves. 
This implies that whatever domain knowledge was accumulated by the RL manager, it was not being used to judge the move recommendations on their merits, as the expert manager does. Instead, relative advantages are ascertained with a more abstract understanding of the qualities of the team members.

\subsection{Are learned relative advantages interpretable?}

It is interesting to test whether the RL manager has learned to distinguish relative advantages between Maia and Leela using human-understandable features, such as those found in the past between humans and older chess programs by \cite{Bratko}.
This would be informative regarding relative advantages of humans and machines, and it might allow more efficient production of synergy using a rule-based manager instead.
To explore this possibility, we formulated a set of human understandable features of the chess board, based on strategic chess concepts. 
These include:
\begin{itemize}
    \item The move count since beginning of the game (ply);
    \item The material points (using traditional point system);
    \item Number of pawn islands (continuous pawn structures);
    \item Concentration of the pieces (average distances between pairs);
    \item Proportion of pieces that are defended;
    \item Number of legal moves available;
    \item Number of attack moves available;
    \item King freedom (number of squares it can move to)
\end{itemize}

Taking a set of board positions in which the RL manager gave preference to the Maia or Leela move, we extracted the above features.
We then compared the mean value for each feature in positions where Maia was given preference, to the mean value when Leela was given preference.
Results are summarized in table \ref{Feature-based explainability}.
We again measure effect size using non-parametric $A_w$ \cite{Li}. 
Effect size is always shown in the direction of lower to higher mean value. 
A value of 0.5 represents no effect, 1 represents highest effect.
We find that effect sizes are all small. 
This implies that no one feature by itself seems to be relied upon by the RL manager to find synergy.

This doesn't mean that the RL manager doesn't implicitly use some combination of these features. 
To test for this, we distilled the transformer based manager into a fully connected deep neural network (20 layers of width 256), which takes the above features as input, and was trained to predict the output of the transformer network.
Using the distilled model as manager, it managed to achieve significant synergy in the symmetric team over the test opening positions, with WDL=0.47. 
This is less than the transformer model, but significantly better than Maia alone (Z=4.77).
This means that a combination of the features is relevant for synergy.

However, deep neural networks combine features in a complex non-linear fashion, making interpretability hard. 
To investigate whether a linear combination of the features might be useful for producing synergy, we used the same RL training regime outlined in section 2.3 above to train a logistic regression model.
The model converged to a manager that chooses the same player every time and so does not achieve synergy.
We conclude that a complex non-linear combination of the features is required to achieve synergy.

\begin{table}
  \caption{Mean Feature Value Differences}
  \label{Feature-based explainability}
  \centering
  \begin{tabular}{lllll}
    \toprule
    Feature                  & Maia Mean  & Leela Mean  & $A_w$ \\
    \midrule
    Ply                      & 50.1       & 39.1        & 0.55  \\      
    Material Points          & 28.2       & 28.7        & 0.52  \\
    Adversary Material Pts.  & 22.7       & 27.7        & 0.59  \\
    Pawn Islands             & 2.5        & 2.6         & 0.51  \\
    Adversary Pawn Islands   & 2.4        & 2.7         & 0.58  \\
    Defended Pieces          & 0.76       & 0.77        & 0.53  \\
    Adversary Defended Pcs.  & 0.61       & 0.72        & 0.57  \\
    Concentration            & 3.2        & 3.0         & 0.55  \\
    Adversary Concentration  & 2.7        & 3.0         & 0.58  \\
    Legal moves              & 35.1       & 33.1        & 0.54  \\
    Adversary Legal Moves    & 26.7       & 32.5        & 0.60  \\
    Attacks                  & 24.6       & 22.8        & 0.55  \\
    Adversary Attacks        & 18.5       & 22.5        & 0.61  \\
    King Freedom             & 3.0        & 2.4         & 0.57  \\
    Adversary King Freedom   & 2.6        & 2.6         & 0.50  \\
    \bottomrule
  \end{tabular}
\end{table}

\section{Addendum: Maia Human Likeness}

While not essential for our conclusions, we assumed in this work that Maia is a reasonable proxy for humans. 
It has over 50\% accuracy in matching human moves (chess has over 30 available moves per turn on average), more than other chess engines \cite{McIlroy}. 
However, we deploy Maia in a team, potentially leading it to out of distribution positions.
To test if this degrades its human likeness, we check for human biases reflected in Maia's playing style, when deployed in a team. 
\cite{Afek} documented "invisible moves", i.e. positions in which humans miss good moves due to bias, such as:

\begin{itemize}
  \item "Geometric bias" - a bias towards advancing over retreating, presumably due to the human vantage point when playing. 
  Humans are also more likely to identify moves in the center of the board than flanking moves on the periphery.
  \item "Didactic bias" - a bias to follow heuristic rules taught in chess clubs. 
  These include trying to control the center, castling and developing major pieces.
  \item "Brutal moves bias" - a bias towards aggressive moves such as capture and giving check, as opposed to subtler moves.
\end{itemize}

To quantify these, we define the following features (all binary):
\begin{itemize}
  \item Whether a move is a backwards move;
  \item Whether a move is a flanking move, defined as a move that originates and ends in the first three or last three files;
  \item The piece being moved (a binary indicator per piece);
  \item Whether the move gives check;
  \item Whether the move is a capture;
  \item Whether the move is a castling.
\end{itemize}
	
\begin{figure}
  \centering
  \includegraphics[width=1.0\linewidth]{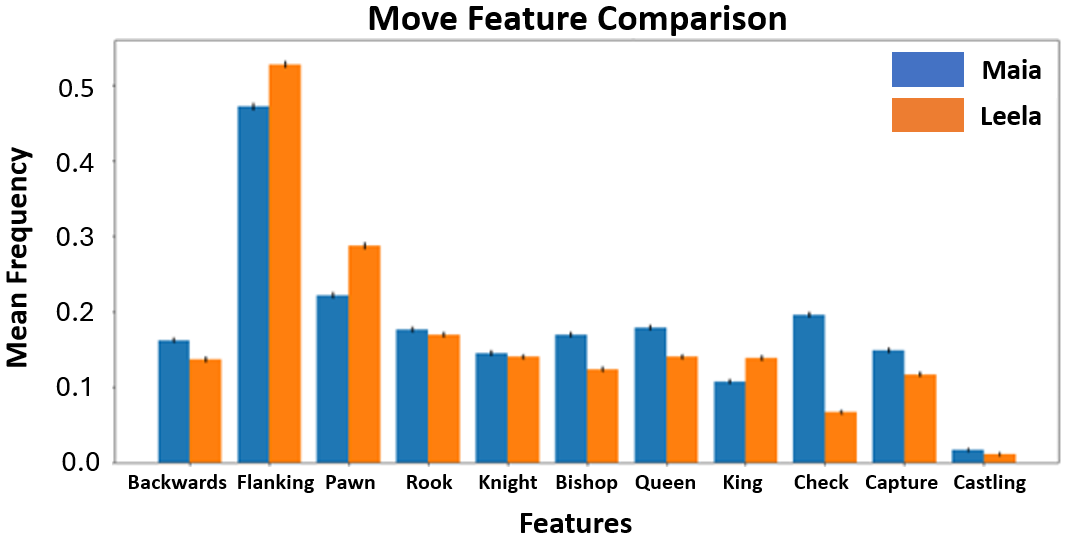}
  \caption{Move feature frequencies compared between Maia and Leela recommendations. Error bars represent SEM.}
  \label{MaiaHumanLike}
  \Description{Comparison of feautures in Maia and Leela moves}
\end{figure}

We generated a set of games with a Maia-Leela team against Stockfish, choosing from their recommendations randomly at each move. 
We then extracted the features of their move recommendations in positions where they disagreed. 
10000 pairs of recommendations were gathered.
The results are shown in figure \ref{MaiaHumanLike}.

Most of the documented human biases we examined seem to manifest in Maia's moves. 
Maia shows a clear bias toward "brutal moves" as shown by her relative preference for captures and giving check. 
Didactic biases are shown by Maia's more frequent castling and the preference to develop major pieces such as knights, bishops and queens. 
Geometric biases are mixed. 
Maia does not show an aversion to backward moves compared with Leela. 
However, Leela is more likely to make a flanking move than Maia.

\section{Related work}

Teams have been known to achieve synergy at various tasks, i.e. better performance than any member of the team. 
This phenomenon has been studied in both human teams and machine ensembles and there is growing work on human-machine teams \cite{Horvitz, Kraus, Shah}. 
In both human-team and machine-ensemble studies, diversity in team composition has been shown to correlate with higher team performance, also known as the "diversity bonus" \cite{Page}. 
This is encouraging for human-machine teams which certainly have cognitive diversity.

However, not all teams achieve synergy. 
In human teams, factors for success include effective communication between team members \cite{Malone}.
This may be complicated in human-machine teams due to lack of explainability of machine models.  
Collective decision-making has also been studied analytically. 
The classic Jury theorem \cite{Condorcet} posits that majority vote leads to correct decisions more often than for any of the voters individually, as long as each voter is correct at least 50\% of the time. 
However, majority vote is not always applicable, for example if there are only two agents, as in our case. 
And it fails when some voters are correct less than 50\% of the time, making voting inappropriate for highly asymmetric teams.

In some cases it is possible to judge the recommendations of team members based on their merit. 
For example, \cite{Osborne} studied boundedly-rational agents as participants in "machine games". 
In machine games, players select from among a set of imperfect "machines" to play on their behalf in a repeated base-game. 
Our work can be viewed as a type of machine game where one base player is human-like and one is an RL trained network.

But choosing based on merit is problematic when there is difficulty verifying the utility of the recommendations. 
\cite{Amir} for example found that human teams failed at solving some types of analytical problems, even when individual team members succeeded. 
This occurred when the correct solution was not easily verifiable and the other team members voted it down. 
We study collective intelligence in a low-verifiability regime (long horizon sequential decision-making).
As we saw, finding a synergetic combination of Maia and Leela moves required a non-trivial process.

Mixture of Experts \cite{MoE} is a successful learning paradigm, that trains a "router" (which we denoted "manager") to select from a set of parallel learners. 
Our work capitalizes on these concepts to model the human-machine team. 
But our setting is bottom-up, taking pretrained frozen policies, and then training only the "manager".

To the best of our knowledge, MoE has not been directly applied to model human-machine teaming. 
Other paradigms have been explored. For example, \cite{Horvitz} advocated end-to-end learning for machine models together with humans. 
Our work by contrast, considers how to leverage pretrained machine models in teams, even though they were not specifically developed for a team setting. 
\cite{Shah} envisioned a division of labor, whereby high-level planning is done by humans, which is then fed to machines for subsequent low-level optimization by the machine. 
By contrast, our work considers when human and machine team members recommend actions in parallel. 
\cite{Debate} proposes a debating mechanism between machines with a human arbitrator for safe AI.
Our work examines how far one can go without information exchange or deliberation between agents. 
\cite{Kraus} developed an agent to assist a human manager in operating multiple robots. 
Our work shares similar motivations.

Chess has been referred to as the "drosophila of AI" \cite{McCarthy}. 
Its simple setup on the one hand, and complex state-space on the other hand, makes it attractive also for studying the human-machine teaming problem.

In chess, \cite{Krakowski} analyzed the relative performance of a set of players in human tournaments, and the same players as centaurs in freestyle tournaments. 
They found no correspondence between the two, i.e. humans don't have to be top players in chess to be effective centaurs. 
They conclude from this that in order to succeed as centaur players, an entirely different set of skills is required than domain knowledge in chess. 
They speculated that these might be technical skills such as data analysis or configuring the machine.

We show that there exists significant potential for synergy in centaur teams by contributing substantive human or humanlike judgement, as with the Maia engine, even if the partner machine is vastly superior.
However, we also show that learning these relative advantages becomes difficult as asymmetry of the team increases.
Furthermore, we find indications that the function of successfully learning to recognize relative advantages between team members, requires at least some domain expertise. 
This might indicate that a human centaur player with no knowledge of chess would probably fail to achieve synergy.

Attempts to characterize human-machine relative advantages in chess are not new. 
Last century, Bratko and Kopec \citep{Bratko} devised a set of tests on different types of positions, and found that while machines excel at solving "tactical" problems, humans are better at "positional" problems, particularly regarding pawn structures.
We conducted a Bratko-Kopec-like experiment for the deep-learning era and found that traditional strategic concepts in chess no longer explain relative advantages (Section 4.3).

More recently, \cite{Wang} trained an adversarial network to beat a superhuman network in the game go, but then showed that the adversarial policy was trivially defeatable by humans. 
That is, they produced a violation of transitivity in game skill between a superhuman policy, an adversarial policy and humans. 
Their work illustrates that even superhuman networks can develop blind spots.
In fact, there are collections of peculiar chess problems that are difficult for superhuman chess programs, even though humans can reason out their solutions \cite{Steingrimsson}.
\cite{Zahavy} considered a method to improve machine performance on these problems by constructing a team of diverse policies. 
Our work leverages human/machine diversity specifically.

Behavioral clones are a popular method for generating scaled synthetic data on human behavior to train collaborative agents \cite{Kantack}.
Here we use them to test for synergy in a human-machine team.

\balance

\section{Conclusions}

In the years ahead, teams will contain not only humans, but also machines of various kinds and proficiencies.
To achieve collective intelligence, these teams will require human intuition and/or automated tools, to guide the level of confidence in the recommendations of team members. 
This work reproduced a simplified version of the centaur phenomenon from freestyle chess in a lab setting, to help investigate how this might be accomplished. 
Specifically, we focused on the function of ascertaining relative advantages between team members and the role of domain knowledge.

\subsection{Contributions}

Firstly, we demonstrate (with the oracle manager) that there is potential for substantial human-machine synergy, which persists even in highly asymmetric teams. 
However, in practice finding the relative advantages that lead to this synergy is non-trivial. 
We propose modeling the human-machine team as a Mixture of Experts to isolate the function of finding relative advantages.
Practically, a MoE model for human-machine teams, could inform team decisions with a "relative advantage score". 
This might be better than using, for example, confidence scores, for which neural networks are not well calibrated \citep{Calibration}, and which are not comparable between different kinds of agents.

We found that a subject matter expert can find a moderate portion of the potential relative advantages and that increasing expertise has quickly diminishing returns for this task.
The limitations of subject matter expertise are probably due to the "curse of knowledge", i.e. the tendency of the expert to overestimate the team members when planning ahead.

We also found that relative advantages between agents can be learned, without communication or explainability of their recommendations, in both symmetric and moderately asymmetric teams. 
This is encouraging for humans teaming with strong but unexplainable machine models. 
Furthermore, we find (by examining the network of the RL manager) that identifying relative advantages requires at least some minimal domain knowledge, even if it is not superior to that of the team members. 
This means that human operators should not forego obtaining domain knowledge, even when machines reach higher levels of proficiency.

Lastly, we saw that finding relative advantages, whether using subject matter expertise or statistical learning, degrades with asymmetry in the team, even while high potential synergy remains.
This is because there are only a small number of high-impact decisions that the inferior team member has to contribute, making the data imbalanced.
We leave the identification of such sparse-but-high-impact decisions as an open challenge.
Note however, that the original centaur teams were not restricted to passive classifications of relative advantages, and likely employed more sophisticated methods.

To reach our conclusions, we made use of a behavioral clone - the Maia network \cite{McIlroy}. 
Our methodology expands the use of behavioral clones, from serving to train collaborative agents, to testing human-machine synergy.
 
\subsection{Limitations and Future Work}

Despite our independent verification of Maia's human likeness, our work stopped short of evaluating teams with real humans.

Our RL trained managers were tailored to the specific team members and adversary.
Learning relative advantages in ad-hoc team compositions with a single model is an open problem which we did not tackle here.

We adopted a simple model in which two agents independently offer recommendations and a manager makes a binary choice between them. 
Future work might examine protocols for deliberation and joint planning between the agents as part of the discovery of relative advantages. 

Ethically, any generic method to improve performance on sequential decision tasks could theoretically be used for both constructive and malicious purposes. 
Generally, however, a focus on human-machine teaming supports a human-centric view of AI development.

Code, data and models are available at:

\url{https://github.com/ReserveJudgement/Centaur-GPT/tree/main}



\begin{acks}
    This work was supported by the Gatsby Charitable Foundation (YL).  Y.L. is the incumbent of the David and Inez Myers Chair in Neural Computation.
\end{acks}



\bibliographystyle{ACM-Reference-Format} 
\bibliography{Centaur}

\ccsdesc[500]{Human-centered computing~Empirical studies in HCI}

\end{document}